%Paper: hep-ph/9302317
%From: HSLA@PHYS.TAMU.EDU
%Date: Fri, 26 Feb 1993 15:38:48 CST

\magnification 1200
\input tables.tex
\overfullrule=0pt

\def\lam{\hbox{$\lambda\kern-6pt^{\_\_}$}}

\def\r{\vbox{\hbox{\raise1.5mm\hbox{$>$}}
\kern-18pt\hbox{\lower1.5mm\hbox{$\sim$}}}}
\def\l{\vbox{\hbox{\raise1.5mm\hbox{$<$}}
\kern-18pt\hbox{\lower1.5mm\hbox{$\sim$}}}}

\def\leaderfill{\leaders\hbox to 1em{\hss.\hss}\hfill}
\def\rs{\vbox{\hbox{\raise1.1mm\hbox{$>$}}
\kern-18pt\hbox{\lower1.1mm\hbox{$\sim$}}}}
\def\ls{\vbox{\hbox{\raise1.1mm\hbox{$<$}}
\kern-18pt\hbox{\lower1.1mm\hbox{$\sim$}}}}

\baselineskip=18pt
\line{\hfill CTP-TAMU-65/92}
\line{\hfill NUB-TH-3055/92}
\line{\hfill hep-ph/9302317}
\bigskip
\centerline{\bf COSMOLOGICAL CONSTRAINTS AND SU(5)}
\centerline{\bf SUPERGRAVITY GRAND UNIFICATION}
\medskip
\centerline{R. Arnowitt}
\centerline{Center for Theoretical Physics, Department of Physics}
\centerline{Texas A\&M University, College Station, TX 77843-4242}
\medskip
\centerline{and}
\medskip
\centerline{Pran Nath}
\centerline{Department of Physics, Northeastern University}
\centerline{Boston, MA 02115}
\bigskip
\centerline{\bf ABSTRACT}
\medskip

\item{} The predictions of SU(5) supergravity models with radiative breaking
constrained by experimental proton decay bounds are discussed.  It is shown
that cosmological
constraints further restrict the parameter space but can be satisfied for a
wide range of parameters.  It is also shown that no serious fine tuning
problems (either at $M_{SUSY}$ or $M_{GUT}$) exist. \bigskip

\line{\bf 1. INTRODUCTION\hfill}

The observation last year [1] that the $SU(3)_C \times
SU(2)_L \times U(1)_Y$ gauge coupling constants, $\alpha_3$, $\alpha_2$ and
$\alpha_1 \equiv (5/3)\alpha_Y$, meet at a common energy scale $\mu = M_G$ if
extended from their measured values at $\mu = M_Z$ by the supersymmetric
renormalization group equations (RGE) with one pair of Higgs doublets, has led
to a number of investigations of other predictions of supersymmetric GUT
models [2-9].  In supergravity grand unification [10] with radiative breaking
of $SU(2) \times U(1)$ [11] (for reviews see [12]), the model depends on four
parameters (aside from the as yet unknown t-quark mass $m_t$):  $m_0$
(universal scalar mass), $m_{1/2}$ (universal gaugino mass), $A_0$ (cubic soft
breaking mass) and \break\hfill tan $\beta \equiv <H_2>/<H_1>$, where $<H_2>$
gives mass to the up quarks and $<H_1>$ to the down quarks and leptons.  Thus
with gauge group $SU(5)$, Refs. [5 and 3] discuss the No-scale model
$A_0=0=m_0$, Ref. [7] the case $B_0=A_0-m_0$ (where $B_0$ is the quadratic
soft breaking mass and can be expressed in terms of tan $\beta$) and Refs.
[2,4,6,7] examine the general parameter space.  Ref. [9] is concerned with the
$O$ (10) model.

The fact that supergravity grand unification introduces only four unknown
parameters (only two more than in the Standard Model itself) to account for
the masses of 32 particles (31 new SUSY particles plus the light Higgs $h$)
implies that there should be considerable correlation between the SUSY masses.
Unfortunately if one imposes only the requirement that radiative breaking
occur and the current experimental bounds on the SUSY masses, the allowed mass
bands are still very broad [2,5,7] and it is difficult to make clear
predictions that can be used to test the theory.  The situation changes
considerably, however, if the model possesses an $SU(5)$-type proton decay:
$p\rightarrow{\bar\nu}K$.  Thus if one assumes no extreme fine tuning of
parameters ($m_0$, $m_{\tilde g} < 1$ TeV where ${\tilde g}$ is the gluino)
and the superheavy Higgs color triplet, which mediates the decay obeys
$M_{H_3}<3M_G$ (which in simple models is the bound that keeps the GUT
couplings perturbative in size), then the parameter space allowed by current
data is still fairly large e.g.:  $m_0~\r~550$ GeV, $m_{\tilde g}~\l~450$ GeV
(i.e. $m_{1/2}~\l~150$ GeV), $1.1 \leq {\rm tan}\beta\leq 4.7$ [2].  However,
one finds a number of remarkable predictions for the SUSY masses [2]:

$$\eqalign{2m_{{\tilde Z}_1}&\simeq m_{{\tilde Z}_2}
\cong m_{{\tilde W}_1}\simeq ({1\over 3} -
{1\over 4})m_{\tilde g}\cr
m_{{\tilde Z}_3}&\cong m_{{\tilde Z}_4} \cong m_{{\tilde W}_2}\cr}\eqno(1)$$

\noindent
where the charginos (${\tilde W}_i$, $i=1,2$) and neutralinos (${\tilde Z}_i$,
$i=1\ldots 4$) are labeled such that $m_i < m_j$ for $i<j$.  In addition one
finds $m_h~\l~110$ GeV and $m_t~\l~180$ GeV.
Further, for $m_t < 140$ GeV, then $m_{{\tilde W}_1}~\l~100$ GeV
 when $m_h~\r~95$ GeV making one of
these particles (and possibly both) observable at LEP200.

The above proton decay constraint is sufficiently powerful to eliminate the
preferred models of Ref. [7] and the No-scale $SU(5)$ model over the entire
parameter space [3].  (For the No-scale case one need not even use the fine
tuning constraint if one imposes the cosmological requirement that the LSP be
electrically neutral.)  Thus No-scale models are viable only if they can
suppress the $p\rightarrow{\bar\nu}K$ decay mode, such as is done in the
flipped $SU(5)$ supergravity model [13].

The purpose of
this letter is to discuss the role of cosmology in a GUT theory which allows
proton decay via dimension five operators.  We shall show that a supergravity
GUT theory which is constrained both by proton decay limit and the
cosmological relic density limit (which avoids overclosing the universe)
allows a wide domain of the parameter space on a reduced four dimensional
manifold (more precisely a five dimensional shell)
in contradiction to the conclusions of a recent analysis on this topic [14].
We also show that the conclusions drawn in Ref. [14] concerning fine tuning
are inaccurate, and there are no serious fine tuning problems either at
$M_{SUSY}$ or $M_G$.

 \medskip
\line{\bf 2.  COSMOLOGICAL CONSTRAINTS\hfill} \smallskip

Recently detailed analyses of the neutralino relic density in N=1 supergravity
unified models have been carried out using the superparticle spectrum
generated by the radiative electroweak symmetry breaking [14-16].  It is found
that the relic density constraint $\Omega_{{\tilde Z}_1}h^2~\l~1$ (where
$\Omega_{{\tilde Z}_1}$ is the ratio of the lightest neutralino mass density
to the critical mass density and $h$ is Hubble constant measured in units of
100 km/sec Mpc) limits additionally the allowed parameter space of the
supergravity models when one imposes the naturalness condition discussed above,
$m_{\tilde g}$, $m_0~\l~1$ TeV.  [If squark and gluino masses in excess of 5
TeV are allowed, the relic density constraint is easily satisfied [15] (due to
the many open channels), as is the p-decay constraint (due to the suppression
from the large SUSY masses in the dressing loop [17]).]

A detailed analysis of the allowed parameter space under the simultaneous
constraints of proton stability [17] and neutralino relic density not
overclosing the universe [18] will be given elsewhere [19].  We give here a
brief discussion.  For an arbitrary point in parameter space one finds
$\Omega_{{\tilde Z}_1}\approx 100$.  As has been pointed out [14-16], however,
the neutralino annihilation rate is significantly enhanced when the
annihilation occurs close (within a few GeV) to the $h$ boson being on shell,
i.e. $2m_{{\tilde Z}_1}\simeq m_h$.  To calculate the ${\tilde Z}_1$ relic
density
near the $h$ pole, we follow the general analysis of Ref. [18] making use of
the cross section for ${\tilde Z}_1 + {\tilde Z}_1 \rightarrow
h^{\ast}\rightarrow f+{\bar f}$ where $f$ is a final state fermion [20].
However, near the pole, it is necessary to take the thermal average of the
rigorous cross section, $<\sigma v>$ ($v$=relative velocity), as discussed in
detail in Ref. [21], rather than use the expansion $\sigma v\cong a + bv^2/6$.
The thermal average can no longer be done analytically, but must be performed
numerically.  (To our knowledge, previous calculations of SUSY relic densities
have not included this important modification.)  From Eq. (1), the condition
that the intermediate $h$ is nearly on-shell
can be viewed as a constraint relating
$m_{\tilde g}$ to $m_h$, and we find strong suppression of $\Omega_{{\tilde
Z}_1}h^2$ over a range of gluino masses \l~5-20 GeV wide.  Thus the inclusion
of the relic density constraint reduces the five dimensional parameter space,
$m_0$, $m_{\tilde g}$, $A_0$, tan $\beta$ and $m_t$ to a four dimensional
submanifold (actually a five dimensional shell $\approx$ 5-20 GeV wide) where
the annihilation of the ${\tilde Z}_1$ is enhanced so that $\Omega_{{\tilde
Z}_1}h^2 < 1$.  If $m_t$ is experimentally determined (as one hopes it soon
will be at the Tevatron) then one will be left with a three dimensional
subspace (more precisely a four dimensional shell) depending on the parameters
$m_0$, $A_0$, and tan $\beta$.

While the relic density constraint reduces the allowed range of the remaining
parameters somewhat, significantly it still leaves available a wide range of
these parameters.  Fig. 1 shows the allowed region in the parameters
$m_{\tilde g}$, $\alpha_H$ (tan $\beta \equiv 1/{\rm tan} \alpha_H$) for a
characteristic example $m_0 = 600$ GeV, $A_t = 0$, $\mu > 0$ where $A_t$ is
the t-quark $A$ parameter at the electroweak scale.
(In Figs. 1 and
2, we have allowed the more conservative bound of $M_{H_3}< 6~M_G$.)  We see
that $m_{\tilde g}$ ranges from 200 GeV to 450 GeV and
$22^{\circ}\leq\alpha_H<41^{\circ}$, much as when only the p-decay constraint
was imposed [2].  (The allowed $m_{\tilde g}$ band is $\approx$ 10 GeV wide.)
Note also that allowed ranges of parameters satisfying both relic density and
proton decay constraints exist for a {\it full range of values of $m_t$ and not
for only ``special'' values of $m_t$} as stated in Ref. [14].  Fig. 2 shows
the dependence of the allowed region on $m_{\tilde g}$ and $A_t$ for $m_0 =
600$ GeV and $\alpha_H = 30^{\circ}$ (tan $\beta = 1.73$).  The $m_{\tilde g}$
band is again 10-20 GeV wide and the range of $A_t$ is similar to
that obtained before [2].
Again, allowed regions in parameter space exist for a range of values of
$m_t$. Fig. [3] shows the importance of using the rigorous analysis of Ref.
[21] rather than the approximation $<\sigma v> = a + b x$ ($x =
kT/m_{{\tilde Z}_1}$).  As one can see from
comparison of the exact and the approximate analysis, the approximate analysis
would introduce large errors.

The above analysis shows that the relic density constraint combined with the
proton decay constraint for arbitrary $m_t$ leads to a wide range of allowed
parameters, in contrast to the analysis in Ref. [14].  Further, the model has
many experimentally testable predictions.  There is, however, an alternate
framework which retains all the predictions of the standard $SU(5)$ model in
terrestrial experiments, and eliminates the constraint of the neutralino relic
density altogether by a sufficiently rapid decay of the neutralinos.  We begin
by recalling that we are dealing with an $N=1$ supergravity theory which is an
effective remnant theory at the scale $M_G$ of some more unified
structure.  As such, it can possess at the scale $M_G$ operators with dim $>
3$ in the superpotential scaled by $M_{P\ell}$ or the compactification mass
$M_C$.  While the effect of these operators on the computation of the
superparticle spectrum via radiative breaking [2-9] would be negligible, they
can significantly affect the cosmology resulting from the model.

As an example, consider the second generation of quarks and leptons and
supplement the particle content of the model by an $SU(5)$, singlet field
$\nu^c$ and an $SO(10)$ singlet field $N$ so that they can be grouped into
nonets of $SU(3)_C \times SU(3)_L \times SU(3)_R$.  We denote these nonets by
the representation $L(1,3,{\bar 3})$, $Q(3,{\bar 3},1)$ and $Q^c({\bar
3},1,3)$ of $[SU(3)]^3$.  Next we extend the superpotential of the theory by
adding the minimal terms $W_s + \lambda [(Tr\Sigma^2)/M^2_C]TrQLQ^c$, where
$W_s$ contains the singlet fields $\nu^c$ and $N$ and generates superheavy
VEVs for them, and $\Sigma$ is the 24 of $SU(5)$ whose VEV, diag$<\Sigma > =
M(2,2,2,-3,-3)$ breaks $SU(5)$ to $SU(3)_C \times SU(2)_L \times U(1)_Y$ at
$M_G$.  The second term contains the factor

$$TrQLQ^c = -DND^c + D\ell^cu^c - D\nu^cd^c - q\ell D^c + qHu^c +
qH'd^c\eqno(2)$$

\noindent
where $D,D^c$ are the superheavy Higgs color triplets, $\ell$ and $q$ are
lepton and quark fields, and $H$ and $H'$ are the two light Higgs doublets.
We note that both $SU(5)$ and $[SU(3)]^3$ can be embedded into $E_6$, pointing
to a possible common origin of the normal and ``Planck slop'' terms from a
more unified structure.

After spontaneous breaking, $W_s$ generates a VEV for $\nu^c$ which
spontaneously produces a violation of R-parity as can be seen from Eq. (2).
This also gives superheavy masses to the $N$ and $\nu^c$ fields.  The $\nu^c$
VEV growth generates a $D-d$ and $D^c-d^c$ mixing.  Diagonalization of the
$D-d$, etc. mass terms [22] gives an effective R-parity violating interaction
in the quark-lepton sector which can decay the lightest neutralino.  Thus
after diagonalization, the R-parity violating interaction determining this
decay is

$${\cal L}_{int} = g({\bar{\tilde Z}}_1\mu_L)({\bar s}^ac^a_L)~;\quad g =
{2e\over m^2_{\tilde\mu}} f_{22}U s_1\eqno(3)$$

\noindent
where $\mu (x)$ is the muon field, $s^a(x), c^a(x)$ ($a =$ color index)
the strange and charm quark fields, $f_{22}$ the $SU(5)$ coupling in $f_{ij}
{\bar H}_x{\bar M}_{iY}M^{XY}_j$ (${\bar H}_x = {\bar 5}$ Higgs field, ${\bar
M}_Y = {\bar 5}$, $M^{XY} = 10$ matter fields, $i,j =$ generation indices),
$U$ is the projection of the ${\tilde Z}_1$ state onto the photino state,
$s_1$
is the $D-d$ mixing parameter given by

$$s_1\cong -y/(m^2_D + y^2)^{1/2}~;\quad y = {<\Sigma^2>\over M^2_C}
<\nu^c>\lambda\eqno(4)$$

\noindent
and $m_D \equiv M_{H_3}$ is the superheavy Higgs color triplet mass.  The
partial lifetime of the ${\tilde Z}_1$ is then

$$\tau ({\tilde Z}_1\rightarrow {\bar c}s\mu^+) \simeq (1\times 10^{-19}{\rm
sec})(f_{22}s_1U)^{-2}({m_{\tilde\mu}\over m_{{\tilde Z}_1}})^4 ({1 {\rm
GeV}\over m_{{\tilde Z}_1}})\eqno(5)$$

\noindent
with $f_{22} = (m_s/M_Z)(e/\sin \alpha_H\sin 2\theta_W)$.  For $<\Sigma > =
<\nu^c> = M_G \cong 10^{16}$ GeV, $M_C = 5 \times 10^{17}$ GeV, $M_D = 3 M_G$,
$\lambda =1$, $U=1$, $m_{\tilde\mu} = 500$ GeV, $m_{{\tilde Z}_1} = 50$ GeV,
one finds $\tau \approx 10^{-4}$sec.  Thus typically the neutralino is very
short lived so that it will not leave any significant cosmological trace.
However, the lifetime of the neutralino is still large enough that it will
decay well outside the detection chamber in collider experiments, so that all
of the characteristic missing $E_T$ signals of supersymmetric particles will
be maintained.
\medskip

\line{\bf 3.  FINE TUNING AT M$_{\rm SUSY}$\hfill}
\smallskip

The problem of fine tuning first arose in non-SUSY GUTs due to the quadratic
divergence of $m_H$, the Higgs mass i.e. $m^2_H = m^2_0 -
b{\tilde\alpha}\Lambda^2$, where $m_0$ is the bare mass, ${\tilde\alpha}$ is a
coupling constant, $\Lambda$ is the cutoff and $b$ is a constant.  Thus if
$m_H = O~(M_Z)$, then one must fine tune $m^2_0$ to 24 decimal places when
$\Lambda = M_G$.  One may formalize this argument [23,14] by defining the
parameter $c \equiv (m^2_0/m^2_H)(\partial m^2_H/\partial m^2_0)$.  Then $c =
m^2_0/m^2_H \cong c{\tilde\alpha}\Lambda^2/m^2_H \simeq 10^{24}$, i.e. log $c$
is the number of fine tuning decimal places.  In supersymmetry the fine tuning
problem resurfaces in the radiative breaking equation [11]

$${1\over 2} M^2_Z = (m^2_{H_1} - m^2_{H_2}{\rm tan}^2\beta )/({\rm
tan}^2\beta -1) - \mu^2~.\eqno(6)$$

\noindent
Here $\mu$ is the $H_1-H_2$ mixing mass, $m_{H_{1,2}}$ are the Higgs masses
which can be expressed in terms of the GUT scale parameters $m_0$, $m_{1/2}$,
$A_0$, $B_0$, $\mu_0$ by the renormalization group equations.  The situation
here is more complex as there are many parameters $a_i = m^2_0$, $\mu^2_0$,
$m^2_{1/2}$ etc.  One may define $c_i \equiv (a_i/M^2_Z)(\partial
M^2_Z/\partial a_i)$ and require $c_i < \Delta_i$ with, say $\Delta_i = 10^2$
as in Ref. [14].  There are however, a number of ambiguities that need to be
addressed.  Thus one can always make transformations on the parameters, $a'_i
= f_i(a_i)$ sending $c_i$ to $c'_i$, increasing or decreasing the value of a
given $c_i$ in this way.  (The $a_i$ in general are complicated functions of
the hidden sector of the theory [12].  Thus which set of functions of $a_i$
are ``fundamental'' and hence to be preferred is unknown at present.)  Also
rescaling all parameters to a single one e.g. $m_{1/2}$ by writing $\xi_0 =
m_0/m_{1/2}$, $\xi_A = A_0/m_{1/2}$ etc. (as done in Ref. [14]) artificially
increases the remaining $c$-parameter as $c_{1/2}$ then equals $\Sigma~c_i$.

In Ref. [14], the choice $a_i = \mu^2$ and $m^2_t$ was made.  The authors then
found only $c_t$ ``too high'' (and then only by a factor of 2-3).  We believe
it is incorrect to use $m_t$ as a fine tuning parameter as (presumably) the
top will shortly be discovered, and then the only correct thing would be to
insert its experimental value.  But even allowing the choice $a = m^2_t$, one
could reduce the value of $c_t$ by a factor of two merely by replacing $M^2_Z$
by $M_Z$ in the definition of $c_t$, which would then satisfy the criteria of
Ref. [14].

The above discussion shows that the fine tuning criteria used in Ref. [14] is
ambiguous up to factors of 2-10.  When one fine tunes to 24 decimal places as
in non-SUSY GUTs, these ambiguities are unimportant.  But if one is talking
about conditions such as $\Delta < 10^2$, no clear conclusions can be drawn.
The only reasonable constraint is the qualitative one used in Ref. [2] that
squark and gluino masses be less than 1 TeV (which is also approximately the
detection upper bound at the LHC and SSC).
\medskip

\line{\bf 4.  FINE TUNING AT M$_{\rm G}$\hfill}
\smallskip

We mention briefly some additional points concerning the problem of fine
tuning at the GUT scale.  There are three theoretically satisfactory methods
of breaking $SU(5)$ to the Standard Model group while maintaining light Higgs
doublets and superheavy Higgs triplets.  The first is that originally proposed
in global SUSY GUT models [24], which requires a fine tuning that, however, is
natural due to the SUSY no renormalization theorems.  These models use a VEV
growth of a 24 representation to break $SU(5)$.  The second is the missing
partner models [25] using 50, ${\overline{50}}$ and 75 representations of
$SU(5)$ where the VEV growth in the 75 breaks $SU(5)$.  The third method makes
use of a global $SU(6)$ symmetry in the GUT sector [26].  The breaking of the
GUT group then makes the Higgs doublets pseudo Goldstone bosons and hence
automatically light.  We view this last method as being more elegant than the
missing partner models in either flipped $SU(5)$ or normal $SU(5)$.

The third method illustrates the fact that fine tuning in the GUT sector may
not be as invidious as other fine tunings.  Thus the physics of the GUT sector
is at present unknown, and some higher symmetry (perhaps from string theory)
may naturally force two coupling constants to be equal, thus keeping the Higgs
doublets light:  this is precisely what happens in the $SU(6)$ model above.

\medskip

\line{\bf 5.  DISCUSSION\hfill}
\smallskip

At present there is no acceptable string model, Calabi-Yau, orbifold or 4-D
construction, which is consistent with the coupling constant unification
analysis [1].  Thus the fact that a GUT model may possibly have string
anticedents appears irrelevant at this point, as it may have to be
significantly modified if and when a viable string model is constructed.  More
relevant is the ``possibility'' that the No-scale model could determine the
soft breaking parameters dynamically.  For the complete No-scale model, where
only the $m_{1/2}$ soft breaking mass is non-zero at the GUT scale, this would
lead to a {\it unique SUSY mass spectrum} at a fixed value of $m_t$.
Implementing this is the most interesting question facing the No-scale model,
since then very likely it could be experimentally determined whether it is
right or wrong.  Barring this theoretical development, one should look for
experimental differences between different supergravity models.  One of these
involves proton decay, where the flipped
$SU(5)$ model suppresses the $p\rightarrow
{\bar\nu}K$ decay [13], while $SU(5)$ supergravity expects it to be seen at
the Super Kamiokande experiment.

This work was supported in part by NSF Grant Nos. PHY-916593 and
PHY-917809.

\noindent
Note added:  After completing this work we received the preprint Ref. [27].
These authors also find a significant region of parameter space simultaneously
satisfying the $SU(5)$ proton decay and cosmological constraints.  They also
reconfirm the mass relations of Eq. (1).
\medskip
\line{\bf FIGURE CAPTIONS\hfill}
\smallskip
\item{Fig. 1.} Region in $m_{\tilde g}-\alpha_H$ parameter space allowed by
the combined proton decay and ${\tilde Z}_1$ relic density constraints for
$m_0 = 600$ GeV, $A_t = 0.0$, $\mu > 0$ (tan $\beta \equiv 1/{\rm
tan}\alpha_H$).
$A_t$ is the t-quark $A$ parameter at the electroweak scale.
The dashed curve is for $m_t = 110$ GeV, solid curve for $m_t
= 125$ GeV, dot-dash for $m_t = 140$ GeV.

\item{Fig. 2.} Region in $m_{\tilde g} - A_t$ parameter space allowed by the
combined proton decay and ${\tilde Z}_1$ relic density constraints for $m_0 =
600$ GeV, $\alpha_H = 30^{\circ}$ (tan $\beta = 1.73$), $\mu > 0$.  Different
curves as in Fig. 1.

\item{Fig. 3.} Contribution to $\Omega_{{\tilde Z}_1}h^2$ from the $h$ pole for
$m_t = 125$ GeV, $m_0 = 600$ GeV, $A_t/m_0 = 0.5$, tan $\beta = 1.73, \mu > 0$.
The solid line is the exact result from thermal averaging over the Higgs pole.
The dashed line is the approximate result when one expands $\sigma v\cong a +
bv^2/6$.

\medskip
\line{\bf REFERENCES\hfill} \smallskip

\item{[1]} P. Langacker, Proc.
PASCOS-90 Symposium, Eds. P. Nath and S. Reucroft (World Scientific,
Singapore, 1990); P. Langacker and M. Luo, Phys. Rev. {\bf D44} (1991) 817; J.
Ellis, S. Kelley and D.V. Nanopoulos, Phys. Lett. {\bf B249} (1990) 441; {\bf
B260} (1991) 131; U. Amaldi, W. de Boer and H. F\"urstenau, Phys. Lett. {\bf
B260} (1991) 447; F. Anselmo, L. Cifarelli, A. Peterman and A. Zichichi, Nuovo
Cim. {\bf 105A} (1992) 581.

\item{[2]} R. Arnowitt and P. Nath, Phys. Rev. Lett. {\bf 69} (1992) 725; P.
Nath and R. Arnowitt, Phys. Lett.
{\bf B289} (1992) 368.

\item{[3]} P. Nath and R. Arnowitt, Phys. Lett. {\bf B287} (1992) 89.

\item{[4]} R. Arnowitt and P. Nath, CTP-TAMU-26/92-NUB-TH-3047-92, to be pub.
Phys. Rev. {\bf D}, November 1, 1992 issue.

\item{[5]} K. Inoue, M. Kawasaki, M. Yamaguchi and T. Yanagida, Phys. Rev.
{\bf D45} (1992) 387; S. Kelley, J. Lopez, H. Pois, D. V. Nanopoulos and K.
Yuan, Phys. Lett. {\bf B273} (1991) 423.

\item{[6]} S. Kelley, J. Lopez, H. Pois, D. V. Nanopoulos and K. Yuan,
CERN-TH-6498/92-CTP-TAMU-16/92.

\item{[7]} G.G. Ross and R.G. Roberts, Nucl. Phys. {\bf B377} (1992) 571.

\item{[8]} M. Drees and M.M. Nojiri, Nucl. Phys. {\bf B369} (1992) 54.

\item{[9]} Q. Shafi and B. Ananthanarayan, BA-91-76; B. Ananthanarayan, G.
Lazarides and Q. Shafi, B-92-29-PRL-TH-92/16.

\item{[10]} A.H. Chamseddine, R. Arnowitt and P. Nath, Phys. Rev. Lett. {\bf
49} (1982) 970.

\item{[11]} L. Iba\~nez, Phys. Lett. {\bf 118B} (1982) 73; J. Ellis, D.V.
Nanopoulos and K. Tamvakis, Phys. Lett. {\bf 121B} (1983) 123;
J. Ellis, J. S. Hagelin, D. V. Nanopoulos and K. Tamvakis, Phys. Lett. {\bf
B125} (1983) 275.
L.
Alvarez-Gaum\'e, J. Polchinski and M. Wise, Nucl. Phys. {\bf B221} (1983) 495;
L. Iba\~nez and C. Lopez, Nucl. Phys. {\bf B233} (1984) 511; L. Iba\~nez, C.
Lopez and C. Mu\~noz, Nucl. Phys. {\bf B256} (1985) 218.

\item{[12]} P. Nath, R. Arnowitt and A.H. Chamseddine, ``Applied N=1
Supergravity'' (World Scientific, Singapore, 1984); H.P. Nilles, Phys. Report
{\bf 110} (1984) 1; H. Haber and G. Kane, Phys. Reports {\bf 117} (1985) 75.

\item{[13]} I. Antoniadis, J. Ellis, J. S. Hagelin and D. V. Nanopoulos, Phys.
Lett. {\bf 231B} (1989) 65.

\item{[14]} J. Lopez, D.V. Nanopoulos and A. Zichichi,
Phys. Lett. {\bf B291} (1992) 255.

\item{[15]} M. Drees and M.M. Nojiri, DESY 92-101/SLAC PUB-5860 (1992).

\item{[16]} S. Kelley, J. Lopez, D. V. Nanopoulos, H. Pois and K. Yuan,
CTP-TAMU-56/92-CERN-TH-6584/92-ACT-16/92-UAHEP 9212.

\item{[17]} R. Arnowitt, A.H. Chamseddine and P. Nath, Phys. Lett. {\bf 156B}
(1985) 215; P. Nath, A.H. Chamseddine and R. Arnowitt, Phys. Rev. {\bf D32}
(1985) 2348.

\item{[18]} H. Goldberg, Phys. Rev. Lett. {\bf 50} (1983) 1490; J. Ellis,
J. S. Hagelin, D. V. Nanopoulos, K. Olive and M. Srednicki, Nucl. Phys. {\bf
B238} (1984) 453.

\item{[19]} P. Nath and R. Arnowitt, NUB-TH-3056/92-CTP-TAMU-66/92.

\item{[20]} J. Ellis, L. Roszkowski and Z. Lalak, Phys. Lett. {\bf B245}
(1990) 545; J. Lopez, D. V. Nanopoulos and K. Yuan, Phys. Lett. {\bf B267}
(1991) 219; Nucl. Phys. {\bf B370} (1992) 445.

\item{[21]} K. Griest and D. Seckel, Phys. Rev. {\bf D43} (1991) 3191; P.
Gondolo and G. Gelmini, Nucl. Phys. {\bf B360} (1991) 145.

\item{[22]} B.A. Campbell, K.A. Olive and D. Reiss, Nucl. Phys. {\bf B296}
(1986) 129; R. Arnowitt and P. Nath, Proc. of Neutrino 88, Ed. J. Schneps, T.
Kafka, W. Mann, and P. Nath, (World Scientific, Singapore, 1988).

\item{[23]} R. Barbieri and Guidice, Nucl. Phys. {\bf B306} (1988) 63.

\item{[24]} S. Dimopoulos and H. Georgi, Nucl. Phys. {\bf B193} (1981) 150; N.
Sakai, Z. Phys. {\bf C11} (1981) 153.

\item{[25]} G. Grinstein, Nucl. Phys. {\bf B206} (1982) 387; H. Georgi, Phys.
Lett. {\bf 108B} (1982) 283; A. Masiero et al, Phys. Lett. {\bf 115B} (1982)
380.

\item{[26]} K. Inoue, A. Kakuto and T. Takano, Prog. Theor. Phys. {\bf 75}
(1986) 664; A. Anselm and A. Johansen, Phys. Lett. {\bf 200B} (1988) 331; A.
Anselm, Sov. Phys. JEPT {\bf 67} (1988) 663; R. Barbieri, G. Dvali and
A. Strumia, CERN-TH-6526/92.

\item{[27]} J. Lopez, D. V. Nanopoulos and H. Pois, CERN-TH-6628/92,
CTP-TAMU-61/92, ACT-19/92.
\bye